\documentclass[letterpaper]{jpconf}
\usepackage{graphicx}
\begin{document}
\title{Where is quantum theory headed?}

\author{Stephen L. Adler}

\address{Institute for Advanced Study,
Einstein Drive, Princeton, NJ 08540, USA}

\ead{adler@ias.edu}

\begin{abstract}
Public talk at the EmQM13 conference opening event on ``The Future of Quantum Mechanics''\end{abstract}

The organizers have asked me to state my views on the direction
of the future development of quantum mechanics.  Will it evolve
within the standard framework, without the addition of new
foundational physics?  Or will the foundations require modification
in an, at least in principle, experimentally detectable way?

First, let's discuss the current status of quantum theory.
Quantum mechanics is our most successful physical theory.  It
underlies our detailed understanding of atomic physics, chemistry,
and nuclear physics, and the many technologies based on this
knowledge.  Additionally, relativistic quantum mechanics is the
basis for the very successful standard model of elementary particles.

However, from its beginnings there have been conceptual problems
associated with the nature of measurement in quantum mechanics.
These can be simply illustrated with the famous Stern-Gerlach
experiment (Fig. 1).
\begin{figure}[b]
\centering
\includegraphics{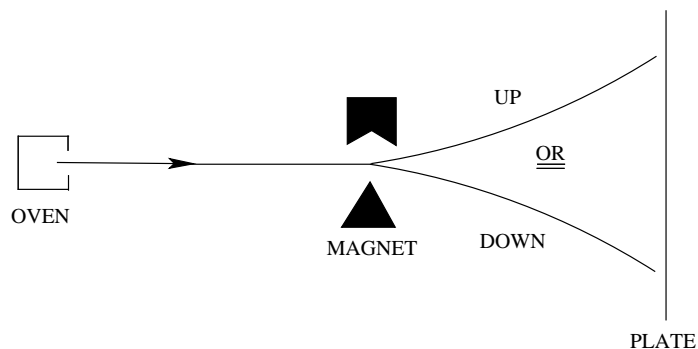}
\caption{Schematic representation of the Stern--Gerlach experiment}
\end{figure}

Silver atoms boiled off from a furnace
are sent through a non-uniform magnetic field, and impinge
on a photographic plate.  Instead of a continuous distribution
of spots, one sees two spots, corresponding to spin up and
spin down relative to the magnetic field axis.  Each atom
goes up OR down, but one cannot predict which in any given
run -- the results of the experiment are probabilistic.
There is a 50\% chance of an atom going up, and a 50\% chance
that it will go down.

From the point of view of  the  Schr\"odinger equation of
quantum theory, this result has no explanation.  In quantum
theory, the state of the particle is described by its wave
function, and the Schr\"odinger equation says that at a
post-measurement final time $T_f$, the wave function is related
to that at a pre-measurement initial time $T_i$, by a deterministic
relation
\begin{eqnarray*}
\Psi(T_f)=&U(T_f,T_i)\Psi(T_i)~~~,\\
U(T_f,T_i)=&e^{iH(T_f-T_i)}~~~,\\
\end{eqnarray*}
with the transition operator $U$ completely specified by the
Hamiltonian $H$.
To explain what is observed, the Schr\"odinger equation must
be supplemented by the reduction postulate and the Born
rule.  These state that the wave function only gives
a description of probabilities when a measurement is made,
with the probabilities for an ``up'' outcome and a
``down'' outcome given by the squares of the coefficients
of the corresponding components in the initial wave
function $\Psi(T_i)$, 
\begin{eqnarray*}
&{\rm Born~Rule~for~Probabilities}\\
\Psi(T_i)=&C_{\rm up} \Psi_{\rm up} +C_{\rm down} \Psi_{\rm down} ~~~,\\
{\rm prob}_{\rm up} =& |C_{\rm up}|^2 ~~~,\\
{\rm prob}_{\rm down} = &|C_{\rm down}|^2 ~~~,\\
&|C_{\rm up}|^2+|C_{\rm down}|^2=1~~~,
\end{eqnarray*}
with the sum of the up and down probabilities equal to one.
The reduction postulate and Born rule are an add-on to the Schr\"odinger
equation. According to the Copenhagen interpretation of quantum mechanics, the
Schr\"odinger equation applies when a microscopic system, the silver atom, is
time-evolving in isolation.  But when the atom interacts with a macroscopic measuring
apparatus, as in the Stern--Gerlach setup, you have to use the reduction postulate
and Born rule.

This situation leads to puzzles that have been debated for over eighty years.
If quantum mechanics describes the whole universe, then why can't one use the
Schr\"odinger equation to describe the system consisting of the silver atom
plus the measuring apparatus?  But we never see a superposition state
of the atom plus apparatus. This is Schr\"odinger's famous  cat paradox. Arrange the
experiment so that an``up'' outcome
triggers a mechanism
that kills the cat, while a ``down'' outcome  keeps the cat alive. Of course we don't do this,
but if we were to do it,  we would always see a live cat
OR a dead one, never a superposition of the two (Fig. 2).   So we have the problem
of definite outcomes:  where does the ``either''--``or''dichotomy arise?
\begin{figure}[t]
\centering
\includegraphics{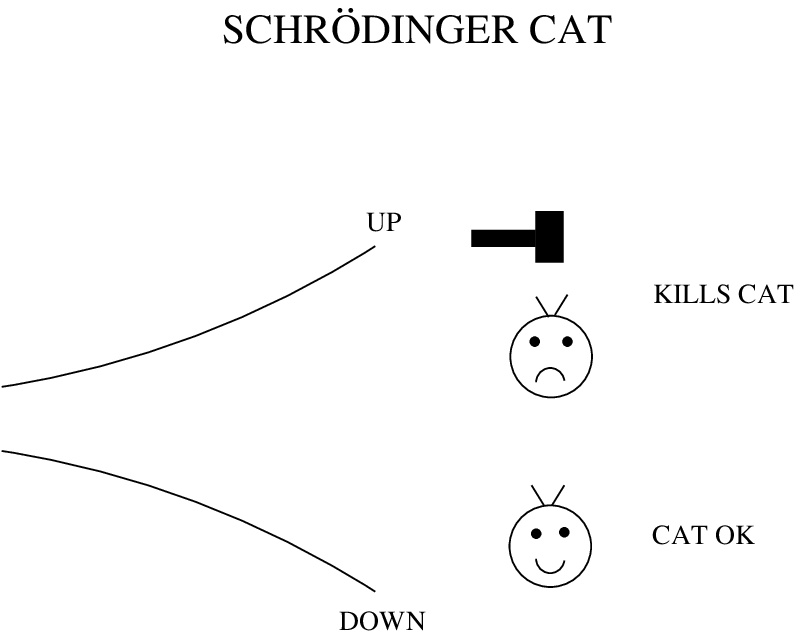}
\caption{ The Stern--Gerlach experiment with a Schr\"odinger cat as the outcome registration}
\end{figure}

A related question is where do the probabilities come from?  Quantum mechanics has
probabilities without a sample space!  An  example of a sample space is a population of
people, 40\% blonde and 60\% brunette.  If you pick a person at random from the
population, there is a probability ${\rm prob}_{\rm blonde}=0.4$ that you will
have a blonde, and ${\rm prob}_{\rm brunette}=0.6$ that you will have a
brunette.  But the population (the sample space) is composed of individuals,
with definite hair coloring -- the probabilities only reflect our ignorance
of details if we make a random pick without looking.
Another example of a sample space,  closer to our
Stern--Gerlach experiment, is a coin toss.  Consider 1000 coin tosses.
If the coin is tossed
without bias, you will find close to 500 heads and 500 tails, corresponding
to ${\rm prob}_{\rm heads}=0.5$ and ${\rm prob}_{\rm tails}=0.5$.  Here the
sample space consists of the 1000 detailed trajectories of the toss, which
your eye cannot follow, but which if analyzed by a very fast computer could
predict which toss would give a head and which a tail (Fig. 3).   Again, the probabilities
are just reflections of our ignorance of the details, but the details are there.
So we have the questions -- are there hidden details underlying the probabilities
in quantum mechanics?  Is there a hidden sample space?

\begin{figure}[b]
\centering
\includegraphics{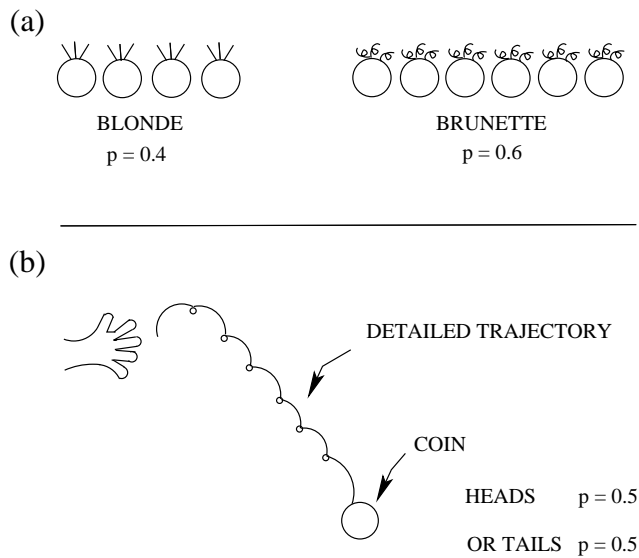}
\caption{Some sample spaces: (a) Populations of individuals with blonde/brunette hair coloring  (b) 
Trajectories in a coin toss}
\end{figure}

Now we come to the question with which I began -- where is future work to
deal with these problems headed?  Two routes proceed within quantum theory:
(i) The first route within quantum theory is to try to change the interpretational rules.  Examples
are the so-called ``many worlds'' interpretation (all possibilities are there, we just only see one), and
the so-called ``histories'' program, which sets up an observer-free generalization
of the Copenhagen rules. (ii) The second route within quantum theory is to say there is a sample space, but
we don't see it.  One example is the proposal of Bohmian trajectories as the sample space underlying the Born rule.
Another example comprises various statistical interpretations, which attribute the probabilistic outcomes
to different internal states of the apparatus and or its environment, with the aim of deriving the Born rule.

None of these routes within orthodox quantum theory has gained general acceptance. Also, none
makes experimental predictions at odds with standard theory, so experimentally they are
not distinguishable.

The other possibility is to modify the foundations of quantum theory. Specifically, one
gets a sample space by postulating additional degrees of freedom -- so called
``hidden variables''.

There are two possibilities for hidden variables -- they can be {\it local}, or
{\it nonlocal}.  Local variables have the property that variables
$V(x_1,t_1)$ and $W(x_2,t_2)$, with $x_1,x_2$ the spatial points and $t_1,t_2$ the times
of occurrence,  cannot influence one another if the distance between
them $|x_1-x_2|$ is greater than the distance $c|t_1-t_2|$ that light can
travel, at velocity $c$,  in the time interval from $t_1$ to $t_2$. Such local variables are called
causally separated.  In quantum theory, causally separated variables have a
commutative multiplication law
\begin{eqnarray*}
V(x_1,t_1)\times W(x_2,t_2)=W(x_2,t_2)\times V(x_1,t_1)~~~.
\end{eqnarray*}
Ordinary numbers obey such a commutative law of multiplication, for example
$AB=BA$ for $A=7$ and $B=11$, whereas in non-commutative multiplication, one would have
$AB \neq BA$.

John Bell's theorem asserts that local hidden variables plus the usual rules for
probabilities imply certain inequalities that are not satisfied by quantum
mechanical systems -- and experiment sees that these inequalities are in
fact violated.  There is much discussion of possible loopholes, but I believe
the result is robust, and that local hidden variables are excluded.

The other possibility is that the hidden variables are non-local:  the hidden
variables can act faster than the speed of light to establish correlations (as long
as no faster than light signaling is possible). The hidden variables can also obey a non-commutative
multiplication law.

For the rest of the talk I'll focus on the possibility of non-local hidden variables --
this is where my research interests lie.  At a phenomenological level, there are
very interesting models for the emergence of probabilities within the usual
wave function formulation of nonrelativistic quantum theory, pioneered by
Ghirardi, Rimini, and Weber in Trieste  and by also by Pearle at Hamilton College
in the U. S., and worked on by many others.
These models postulate that space is filled with a very low level noise
with a coupling to matter proportional to the imaginary
unit $i$, rather than with a real-valued coupling (more technically,
they couple through an  anti-Hermitian Hamiltonian term).  For example, there
could be  a small,
rapidly fluctuating contribution to the gravitational potential or ${g}_{00}$ metric component
proportional to the imaginary unit $i$.  If such a theory obeys two
general properties, (1) the total probability of a particle
being present remains one for all times (that is, the wave function normalization is preserved), and
(2) there is no faster than light signaling, then the extra terms
in the Schr\"odinger equation  equation must have a special structure.  This special
structure allows one to {\it prove} definite outcomes obeying the Born rule!

In these models, for each repetition of the Stern--Gerlach experiment, the noise variable takes
different values.  For a large apparatus, these have a measurable
effect, whereas for an atom not interacting with an apparatus, the
effect is not measurable.  The noise leads to different outcomes
for different runs, with  probabilities  given
by the Born rule.  The different noises for different runs of the experiment are
analogous, in the coin toss example I gave earlier, to different details of the tumbling coin
trajectories for the different coin tosses (Fig. 4).

\begin{figure}[t]
\centering
\includegraphics{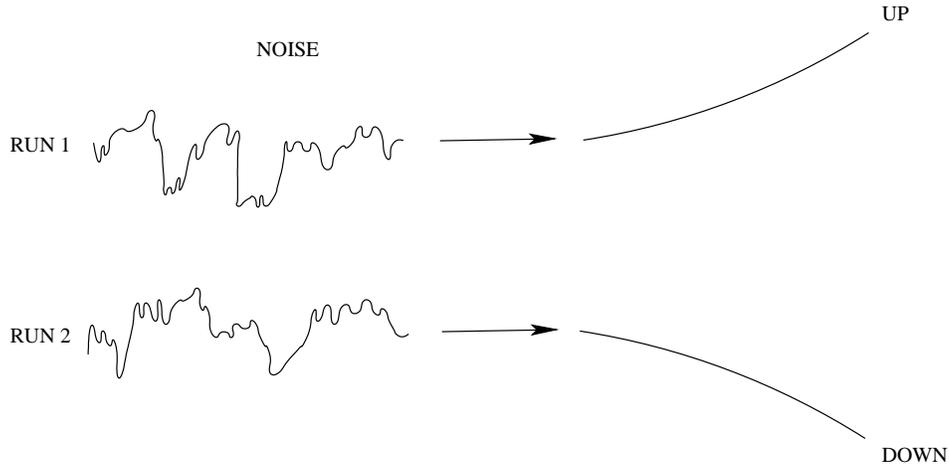}
\caption{Different noise histories, in objective reduction models, can explain
``up'' and ``down'' registrations in the Stern--Gerlach experiment}
\end{figure}

I've worked on phenomenological reduction models, but my main long term interest
has been at the foundational level.  I am trying to make an analogy between
quantum mechanics emerging from a possible pre-quantum theory, and the known
fact that thermodynamics emerges
from the laws of  statistical mechanics.  Thermodynamics --
the science of heat and work (Fig. 5) -- reflects averaged properties of huge
numbers of atoms. It is a complete, consistent system by itself, and remarkably was
discovered in the 19th century {\it before} the existence of atoms was
established.  But from statistical mechanics -- the laws of motion of
large systems of atoms, one can deduce the laws of thermodynamics, together
with details of  fluctuation corrections to thermodynamics, so called Brownian
motion corrections.  (The figure shows the random walk trajectory of a pollen grain
being bombarded by molecules in thermal motion.)

\begin{figure}[t]
\centering
\includegraphics{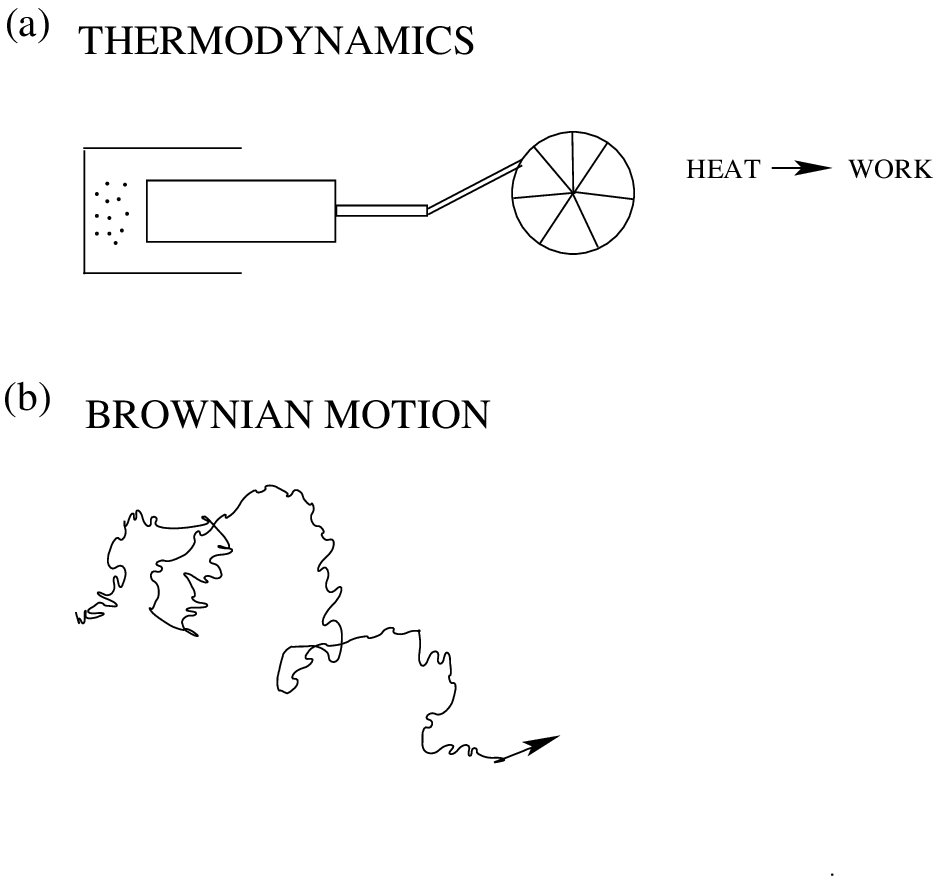}
\caption{(a) Thermodynamics: the science of heat and work  (b)  Brownian motion of a pollen grain
being bombarded by molecules in the  liquid in which it is suspended}
\end{figure}

My suggestion, in articles with collaborators and a small book that I wrote in 2004 \cite{<adlerbook>} , is
a theory that I call  ``trace dynamics''.  It is  a classical-like system
of non-commutating variables -- even distant systems in the
universe are interacting instantaneously with us.  One can make sense
 of the mathematics of non-commuting variables by using the cyclic property
of a mathematical  operation called the Trace: $ {\rm Trace} ABCD..FG
={\rm Trace} GABCD..F={\rm Trace} FGBCD...$   This is a very powerful tool.
One can use it to set up a system of equations analogous to classical mechanics, and
to do statistical averaging.  Getting a little technical now, for those in the
audience familiar with quantum theory and statistical physics, what distinguishes
trace dynamics from ordinary classical mechanics is the existence of a generic
conserved quantity in addition to the energy and momentum.   This quantity is
operator-valued, and has the form
\begin{eqnarray*}
&{\rm Conserved~operator~ in~ trace~ dynamics}=\\
&\sum_{\rm bosons}[ q_{\rm boson}~ p_{\rm boson}-p_{\rm boson} ~q_{\rm boson}]\\
-&\sum_{\rm fermions}[ q_{\rm fermion}~ p_{\rm fermon}+p_{\rm fermion} ~q_{\rm fermion}]
~~~,\\
\end{eqnarray*}
with the $q$s the canonical coordinates and the $p$s the canonical momenta.  
This is reminiscent in structure to the canonical commutation and anti-commutation relations
of quantum theory.  Just as energy in statistical mechanics is equally partitioned
between the various degrees of freedom, one might expect this conserved operator,
in a statistical mechanical treatment of trace dynamics, to also be equi-partitioned,
giving the starting point for quantum theory.

My conjectures thus are:
statistical averages in trace dynamics give the Schr\"odinger equation and operator algebra of quantum theory,
while Brownian motion corrections  give the low level noise on which phenomenological reduction
models are based.  I talked about this program in my keynote address at the Vienna conference
two years ago.  Currently I am working on incorporating gravity into trace dynamics \cite{<adlerpaper>} ,
and that is what I will talk about tomorrow.  My approach to an emergent
quantum theory is  still a work in progress -- there is  much yet to be done!

\ack
I wish to thank Gerhard Gr\"ossing for organizing the conference  EmQM13 on Emergent Quantum 
Mechanics, the Fetzer Franklin fund for its financial support, and Susan Higgins for redrawing my 
figure sketches for publication.

\end{document}